\documentclass[aps,prx,reprint,amsmath,amssymb,superscriptaddress,showpacs,longbibliography]{revtex4-1}
\usepackage{bm}
\usepackage{graphicx}
\usepackage{color}
\usepackage[dvipsnames]{xcolor}
\usepackage{dcolumn}
\usepackage{hyperref}
\usepackage{float}
\usepackage{mhchem}

\begin{document}

\title{Frustrated Magnetism in Mott Insulating (V$_{1-x}$Cr$_x$)$_2$O$_3$}

\author{J. C. Leiner}
\email[]{Corresponding author: jleiner@alumni.nd.edu}
\affiliation{Center for Correlated Electron Systems, Institute for Basic Science (IBS), Seoul 08826, Korea}
\affiliation{Department of Physics and Astronomy, Seoul National University, Seoul 08826, Korea}
\affiliation{Neutron Scattering Division, Oak Ridge National Laboratory, Oak Ridge, TN 37831 USA}

\author{H. O. Jeschke}
\affiliation{Research Institute for Interdisciplinary Science, Okayama University, Okayama 700-8530, Japan}

\author{R. Valent{\'\i}}
\affiliation{Institut f{\"u}r Theoretische Physik, Goethe-Universit{\"a}t Frankfurt, Max-von-Laue-Stra{\ss}e 1, 60438 Frankfurt am Main, Germany}

\author{S. Zhang}
\affiliation{Institute for Quantum Matter and Department of Physics and Astronomy, The Johns Hopkins University, Baltimore, Maryland 21218 USA}

\author{A. T. Savici}
\author{J. Y. Y. Lin}
\author{M. B. Stone}
\author{M. D. Lumsden }
\affiliation{Neutron Scattering Division, Oak Ridge National Laboratory, Oak Ridge, TN 37831 USA}

\author{Jiawang Hong}
\affiliation{School of Aerospace Engineering, and Institute of Advanced Structure Technology, Beijing Institute of Technology, Beijing 100081, China}
\affiliation{Materials Science and Technology Division, Oak Ridge National Laboratory, Oak Ridge, TN 37831 USA}

\author{O. Delaire}
\affiliation{Department of Mechanical Engineering and Materials Science and Department of Physics, Duke University, Durham, North Carolina 27708 USA}
\affiliation{Materials Science and Technology Division, Oak Ridge National Laboratory, Oak Ridge, TN 37831 USA}

\author{Wei Bao}
\affiliation{Department of Physics, Renmin University of China, Beijing 100872, China}

\author{C. L. Broholm}
\affiliation{Institute for Quantum Matter and Department of Physics and Astronomy, The Johns Hopkins University, Baltimore, Maryland 21218 USA}
\affiliation{Neutron Scattering Division, Oak Ridge National Laboratory, Oak Ridge, TN 37831 USA}

\date{\today}

\begin{abstract}
V$_2$O$_3$ famously features all four combinations of paramagnetic vs antiferromagnetic, and metallic vs insulating states of matter in response to \%-level doping, pressure in the GPa range, and temperature below 300~K. Using time-of-flight neutron spectroscopy combined with density functional theory calculations of magnetic interactions, we have mapped and analyzed the inelastic magnetic neutron scattering cross section over a wide range of energy and momentum transfer in the chromium stabilized antiferromagnetic and paramagnetic insulating phases (AFI~\&~PI).  Our results reveal an important magnetic frustration and degeneracy of the PI phase which is relieved by the rhombohedral to monoclinic transition at $T_N=185$~K. This leads to the recognition that magnetic frustration is an inherent property of the paramagnetic phase in $\rm (V_{1-x}Cr_x)_2O_3$ and plays a key role in suppressing the magnetic long range ordering temperature and exposing a large phase space for the paramagnetic Mott metal-insulator transition to occur. 
 \end{abstract}
\maketitle

\section{Introduction}
\begin{figure}
\includegraphics[width=1.00\columnwidth,clip]{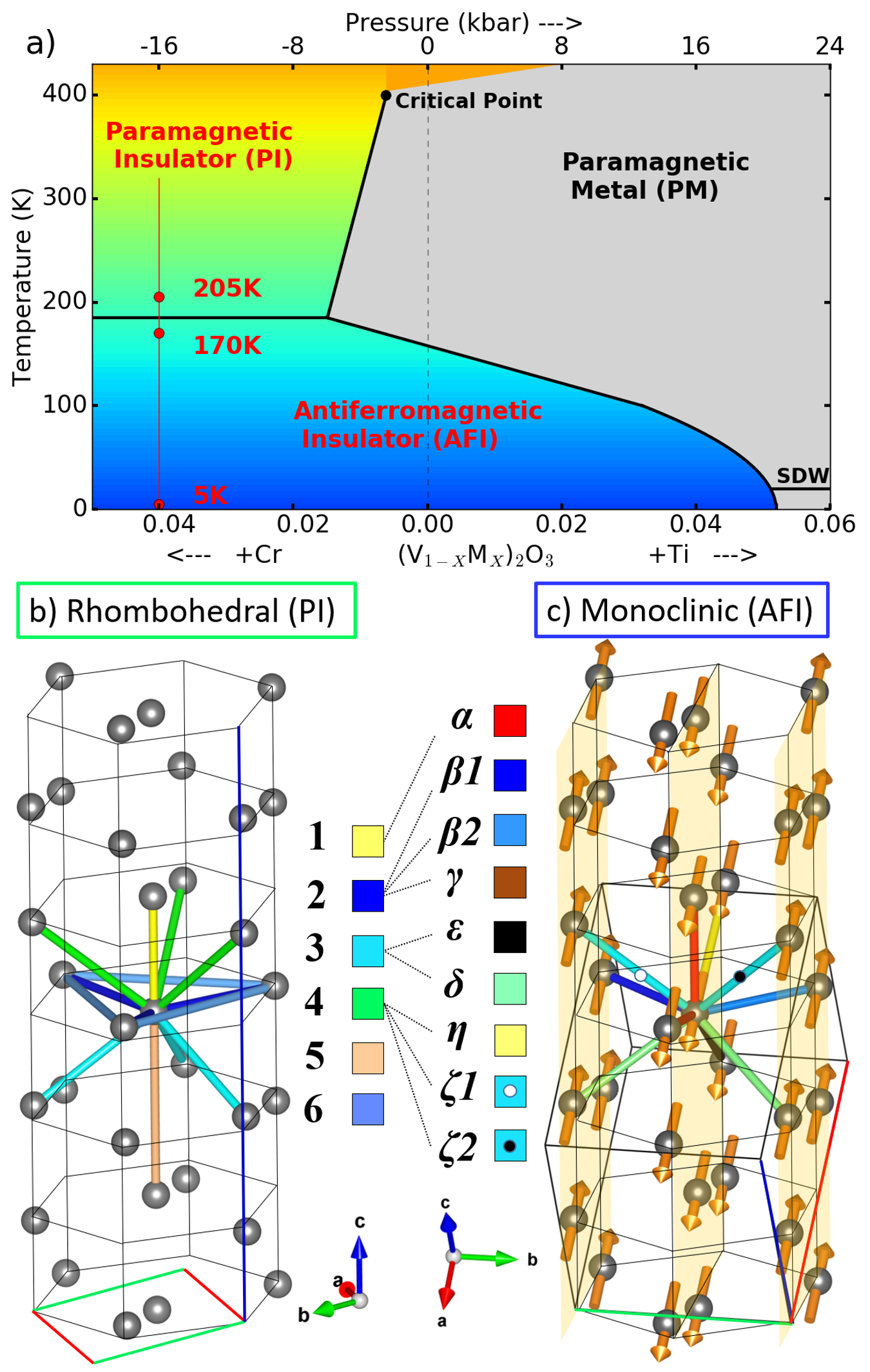}
\caption{ \label{Phase_diagram} (a) $\rm V_2O_3$ phase diagram showing the extent of the AFI, PI, and PM phases \cite{McWhan_1973} as a function of both chemical substitution (bottom axis) and the equivalent applied pressure (top axis). (b) The rhombohedral structure of the PI phase, with the color labeled  exchange interactions (1-6) determined in the paper. Black hexagons outline the pseudo-hexagonal (corundum) unit cell. (c) The distorted crystal structure (monoclinic unit cell shown) and magnetic structure in the AFI phase \cite{Moon_1970}, where the magnetic moments are ferromagnetically aligned in monoclinic a+c planes, which are then stacked antiferromagnetically along the monoclinic b axis. Greek indices indicate the updated nearest neighbor bond configuration resulting from the monoclinic distortion. Oxygen atoms are omitted for clarity. }
\end{figure}

Metal to insulator transitions come in many guises, but for transition metal oxides they generally involve magnetic or structural symmetry breaking. $\rm V_2O_3$ however, offers a singular counter-example. As a function of chromium doping, $\rm (V_{1-x}Cr_x)_2O_3$ undergoes a \emph{paramagnetic} metal to insulator transition that is accompanied by a volume expansion but no global magneto-structural symmetry breaking (Fig.~\ref{Phase_diagram}(a)) \cite{Leonov_2015,PhysRevLett.97.195502,frandsen2016volume}. In this paper we identify \emph{magnetic frustration in paramagnetic insulating} (PI) $\rm (V_{1-x}Cr_x)_2O_3$ as a rare characteristic that contributes to expose the available paramagnetic phase space for the Mott metal-insulator transition~\cite{lechermann2018uncovering} by suppressing long range magnetic order. Furthermore, we show that, in contrast with the PI phase, the magnetic interactions of the monoclinic antiferromagnetic insulating (AFI) phase are all simultaneously satisfied in the ordered state. The magneto-structural transition from the PI to AFI (Fig.~\ref{Phase_diagram}(a)) thus lifts degeneracies associated with frustrated interactions in a spin-Peierls-like transition \cite{spinpeierlstheory} as in other frustrated magnets such as $\rm ZnCr_2O_4$~\cite{ZCO_PRL} and $\rm ZnV_2O_4$~\cite{znv2o4,jpsj_review_2010}.

Our conclusions are based on a careful examination of magnetic interactions in (V$_{0.96}$Cr$_{0.04}$)$_2$O$_3$ through inelastic neutron scattering (INS) and density functional theory (DFT) total energy calculations. The low $T$ monoclinic antiferromagnetic insulator has conventional spin-wave excitations that allow us to determine the non-frustrated exchange Hamiltonian by measuring and analyzing the spin wave dispersion relation. The agreement between the experimentally extracted Hamiltonian and the one calculated by DFT establish DFT as a valid method to determine the dominant exchange constants in $\rm V_2O_3$. We then perform DFT calculations to discover the key exchange constants in the PI and show, using a Gaussian approximation, that these are consistent with measured diffuse magnetic neutron scattering. The PI is a strongly frustrated quasi-two-dimensional honeycomb antiferromagnet with a nominal critical temperature below 10~K. That phase transition is, however, preempted by a spin-Peierls-like magneto-structural transition at 185~K.

Our observations suggest that the large region of paramagnetism in the phase diagram of $\rm V_2O_3$ is made possible by magnetic frustration which deeply suppresses magnetic order in the Mott insulator. This reinforces the importance of exploring metallization of frustrated magnets through doping \cite{PhysRevX.6.041007, mazin2014theoretical} and pressure \cite{kanoda_mott,PhysRevB.83.035106,RevModPhys.89.025003}. 

\subsection{Summary of prior work on $\rm V_2O_3$}
While our work reveals a previously unappreciated aspect of V$_2$O$_3$, it is by no means the first effort to understand spin interactions in this material \cite{McWhan_1969,PhysRevB.2.3734,McWhan_1973,ueda1980phase,PhysRevB.22.2626,PhysRevLett.67.3440,yethiraj1990pure,limelette2003universality,PhysRevB.92.075121}. In a comprehensive work in 1978, Castellani {\it et al.} \cite{Castellani_1,Castellani_2,Castellani_3} invoked the formation of vanadium dimers and e$_g$ orbital order to explain the measured antiferromagnetic (AFM) structure \cite{Moon_1970} in the AFI phase. Many other theoretical studies of V$_2$O$_3$ used simpler one-orbital models which neglect the orbital degrees of freedom and assume the same kind of AFM correlations for all phases of V$_2$O$_3$ \cite{PhysRevLett.75.105}. In 1993, a different kind of AFM order was found in the metallic phase of V$_{2-y}$O$_3$ \cite{PhysRevLett.71.766,PhysRevLett.101.096406} stabilized to low temperatures through vanadium deficiency. It was further shown through INS that the phase transition to the AFI phase, from either the metallic or the insulating paramagnetic phases, abruptly shifts the critical magnetic wave vector from (10$\ell$) to $(\frac{1}{2}\frac{1}{2}0)$ \cite{aeppli1995magnetic,Bao_Broholm_PRL}. This is inconsistent with predictions from one-orbital theories, and indicates the transition is not a conventional order-to-disorder type of magnetic transition. This behavior seemed to be naturally explained within the coupled spin-orbital model of Rice {\it et al.} \cite{rice1995orbital}, as an orbital ordering transition with accompanying antiferromagnetism. The symmetric SU(4) spin-orbital model has since attracted considerable theoretical interest \cite{PhysRevLett.81.3527,PhysRevB.91.155125,PhysRevB.60.6584}.

Subsequent resonant x-ray scattering experiments discovered a new type of Bragg peak \cite{PhysRevLett.82.4719}, which was explained as the order parameter of the e$_g$ orbital ordering \cite{PhysRevLett.80.3400,PhysRevB.65.054413}. Polarization-dependent x-ray absorption spectroscopy (XAS) measurements \cite{PhysRevB.61.11506} reached the unexpected conclusion that there is an $S$~=~1 state at each V site, with dominant orbital occupation in e$_g^\pi$ and a small admixture of the a$_{1g}$ orbital. This was supported by LDA+U \cite{Ezhov_1999} and LDA+DMFT \cite{PhysRevLett.86.5345,hansmann2013mott} calculations with selected values for the Hubbard and the Hund's coupling energies. The calculated magnetic structures in LDA+U \cite{Ezhov_1999}, determined without orbital degeneracy or orbital ordering being necessary, were found to be consistent with experiments. Alternatively, an $S$~=~1 model with orbital degeneracy was suggested \cite{Mila_PRL_2000,Shiina_PRB_2001,PhysRevLett.86.5743} in which stable magnetic and orbital structures were systematically analyzed and anomalous features of the AFI transition were qualitatively explained. In this picture, the phase transition from PI to AFI was interpreted as being fundamentally a structural and orbital occupational ordering transition \cite{Quantum_Melting_PRL} that brings the spin system deeply into a long range ordered state without the usual critical regime associated with a growing spin correlation length.

In recent years, ARPES measurements \cite{PhysRevLett.117.166401} and fully charge self-consistent LDA+DMFT calculations \cite{Grieger_2014,Deng_2014,Leonov_2015,lechermann2018uncovering} indicate a much weaker orbital polarization {than initially suggested \cite{PhysRevLett.86.5345,hansmann2013mott}} that may influence the nature of the paramagnetic metal-to-insulator transition (MIT). Furthermore, the authors of Ref. \cite{lechermann2018uncovering} found that changes in the V$_2$O$_3$ paramagnetic phase diagram are driven by defect-induced local symmetry breakings resulting from different couplings of Cr and Ti dopants to the host system. These results suggest local distortions which lift orbital degeneracy play an important role in the description of the insulating phase in Cr-doped V$_2$O$_3$.

\subsection{Outline}
In view of ongoing efforts to understand the insulating phases in V$_2$O$_3$, it is important to provide an experimental anchor with reliable values of magnetic exchange interactions that limit the parameter space of $S$~=~1 theories. This is possible in the AFI phase by measuring spin wave dispersion relations and comparing those to linear spin-wave theory (LSWT). The acoustic branch of spin waves was previously measured using INS near the magnetic zone center for V$_2$O$_3$ and (V$_{0.96}$Cr$_{0.04}$)$_2$O$_3$ in the AFI phase \cite{Word_1981_PRB,PhysRevB.36.8675,Bao_PRB_1998}. The limited range of those data, however, is insufficient to determine the many distinct exchange interactions of the monoclinic phase \cite{Bao_PRB_1998}.

Here we report INS measurements of both acoustic and optic branches of spin waves in the AFI phase and incoherent magnetic excitations in the PI phase throughout the Brillouin zone for (V$_{0.96}$Cr$_{0.04}$)$_2$O$_3$. At room temperature V$_2$O$_3$ has the trigonal (corundum) structure with space group R$\overline{3}$c (No.~167) while the space group of the low-temperature monoclinic structure is $I$~2/a (No.~15) with 8 vanadium ions per unit cell. Our new measurements and new ab initio DFT calculation methods for (V$_{0.96}$Cr$_{0.04}$)$_2$O$_3$ allow for the accurate determination of all magnetic exchange interactions in the monoclinic long range ordered AFM state of vanadium sesquioxide. We find moderate ferromagnetic nearest neighbor (nn) and dominant AFM next-nearest neighbor (nnn) interactions in an unfrustrated configuration along the zigzag V chains in the AFI phase (Fig.~\ref{Phase_diagram}(c)). Having established the effectiveness of our DFT calculation methods in the AFI phase, we apply them to determine an interacting spin Hamiltonian for the PI phase. A Gaussian approximation applied to this model accounts in detail for our measurements of the short range spin correlations in the PI. This leads to the conclusion that the lack of spin order for temperatures down to 185 K in the PI phase is a consequence of frustrated interactions on the puckered honeycomb lattices that make up the corundum structure. \\

\section{Experimental Methods}
Single crystal samples of (V$_{0.96}$Cr$_{0.04}$)$_2$O$_3$ ($T_N=185$~K) were grown using a skull melter \cite{HARRISON1980571}. To increase the sensitivity of our experiment, four single crystals were co-aligned for a total mass of 17 g. In the PI phase at $T=296$~K, the pseudo-hexagonal lattice constants are $a$~=~$b$~=~4.94~$\mathrm{\AA}$ and $c$~=~14.01~$\mathrm{\AA}$. In the AFI phase at $T=77$~K, the monoclinic lattice parameters are $a$~=~7.28~$\mathrm{\AA}$, $b$~=~4.99~$\mathrm{\AA}$, $c$~=~5.54~$\mathrm{\AA}$, ($\alpha$,$\beta$,$\gamma$~=~[90$^{\circ}$,~96.75$^{\circ}$,~90$^{\circ}$]) consistent with previous findings \cite{PhysRevB.2.3734,PhysRevB.2.3771,PhysRevB.2.3751}. 

We use both the primitive monoclinic unit cell and the pseudo-hexagonal unit cell to label reciprocal space (Fig. \ref{Phase_diagram}). The AFM wave vector \mbox{($\frac{1}{2}$ $\frac{1}{2}$ 0)$_{H}$} in the pseudo-hexagonal unit cell becomes \mbox{(0~1~0)$_{M}$} in the monoclinic unit cell. The conversion between pseudo-hexagonal and monoclinic coordinates is approximately given by:
\[
\left( \begin{array}{c}
H_{M} \\
K_{M} \\
L_{M} \end{array} \right)=\left( \begin{array}{ccc}
\frac{2}{3} & -\frac{2}{3} & -\frac{1}{3} \\
1 &  \,\,\,1 & 0 \\
\frac{1}{3} & -\frac{1}{3} & \frac{1}{3} \end{array} \right)
\left( \begin{array}{c}
H_{H} \\
K_{H} \\
L_{H} \end{array} \right) 
\]

Time-of-flight INS measurements were performed using the SEQUOIA fine-resolution Fermi-chopper spectrometer at the Spallation Neutron Source at ORNL \cite{granroth2010sequoia,stone_review}. (Initial measurements were performed at the MARI multidetector chopper spectrometer at the pulsed spallation source at ISIS, UK). With this modern pulsed-neutron spectrometer, we were able to fully map the excitation spectrum of (V$_{0.96}$Cr$_{0.04}$)$_2$O$_3$ at three key temperatures indicated in Fig. \ref{Phase_diagram}(a). The crystal array was aligned so that the (-$L$ $K$ $L)_{M}$ plane was horizontal. The crystal assembly was rotated through 180$^{\circ}$ in 2$^{\circ}$ steps about the vertical ($H$ 0 $H/2)_{M}$ direction to access a volume of momentum space during the experiment. To balance energy resolution with ${\bf Q}-$space coverage, incident energies of \mbox{50 meV} and \mbox{100 meV} were used to collect data at (5~K, 170~K, 205~K) and (5~K, 205~K, 320~K) respectively. 

The magnetic neutron scattering cross-section for momentum transfer $\textbf{Q}=\textbf{k}_i-\textbf{k}_f$ and energy transfer $\hbar\omega=E_i-E_f$ is given by: 
\begin{eqnarray}
\frac{d^2\sigma}{d\Omega dE_f} &=& N\frac{k_f}{k_i}r^2_0  \, e^{-2W(Q)}|\frac{g}{2}f(\textbf{Q})|^2\nonumber \\
&&\times \sum_{\alpha\beta}(\delta_{\alpha\beta}-\hat{Q}_{\alpha}\hat{Q}_{\beta}){\cal S}^{\alpha\beta}(\textbf{Q},\omega).
\label{xsec}
\end{eqnarray}
where $N$ is the number of hexagonal unit cells, $r_0 = 0.539\times 10^{-12}$ cm, $f(\textbf{Q})$ is the magnetic form factor for the V$^{3+}$ ion, and $e^{-2W(Q)}$ is the Debye-Waller factor. The dynamic spin correlation function (or dynamical structure factor) is given by
\begin{eqnarray}
{\cal S}^{\alpha\beta}(\textbf{Q},\omega) &=& \frac{1}{2\pi\hbar}\int dt \, e^{-i\omega t}\frac{1}{N}\sum_{\textbf{R},\textbf{R}^\prime} 
e^{i\textbf{Q}\cdot (\textbf{R}-\textbf{R}^\prime)}\nonumber \\
&&\times\langle S_{\textbf{R}}^{\alpha}(0) S_{\textbf{R}^\prime}^{\beta}(t) \rangle .
\label{dyncor}
\end{eqnarray}
To enhance statistical quality, we employed the rotational symmetry operations of the PI phase to project the data into an irreducible wedge. In addition, the Mantid software suite \cite{arnold2014mantid} was used to subtract an incoherent background (see Appendix~\ref{append_background}). The intensity data were normalized as described in Appendix~\ref{phonorm} so that we report the measured scattering cross section in absolute units as 
\begin{equation}
{\cal I}({\bf Q},\omega)=\frac{k_i}{k_f}\frac{1}{N}\frac{d^2\sigma}{d\Omega dE_f}
\end{equation}  

\begin{figure*}[!t]
	\includegraphics[width=1.0\textwidth,clip]{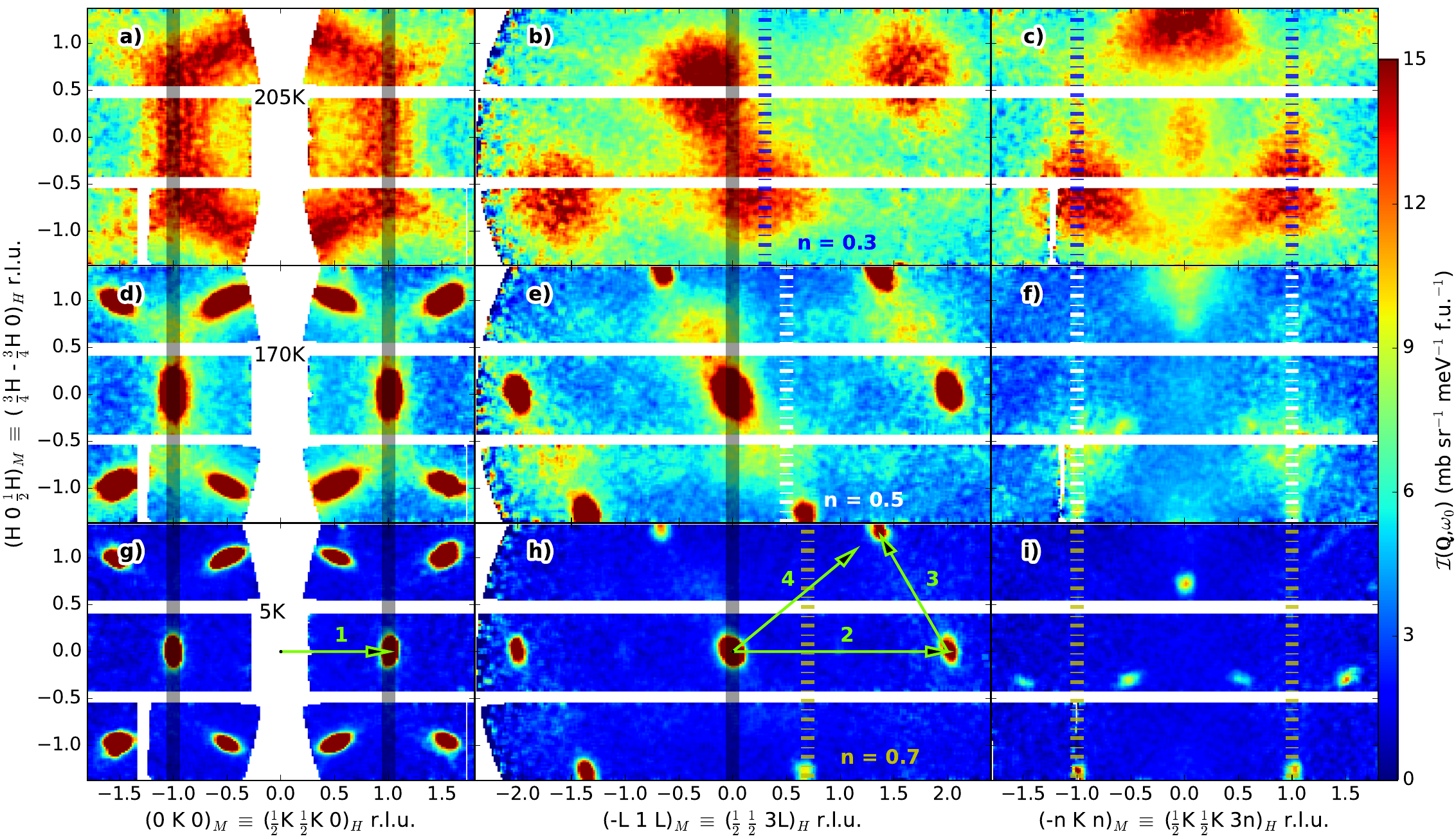}
	\caption{ \label{CES_plots} Constant energy slices of INS data ($E_i=50$~meV) for (V$_{0.96}$Cr$_{0.04}$)$_2$O$_3$; energy transfer averaged from $5~{\rm meV}<\hbar\omega_0<10~{\rm meV}$. Each horizontal row shows ${\cal I}({\bf Q},\omega_0)$ for the three temperatures marked along the phase diagram in Fig. \ref{Phase_diagram}. The left column (a), (d), and (g) shows ${\cal I}({\bf Q},\omega_0)$ in the pseudo-hexagonal basal plane at the origin $(L=0)$. The center column (b), (e), and (h) covers a plane perpendicular to that of the left column overlapping at the light gray solid lines. The right column shows pseudo-hexagonal planes for $L_H=3n$ where $n$~=~0.3, 0.5, and 0.7 for frames (c), (f), and (i) respectively as indicated in the center column by the colored dashed lines. Green arrows indicate the 4 high-symmetry directions used in Fig. \ref{Dispersions}.} 
\end{figure*}

\section{Results}

\subsection{Overview of ${\cal S}({\bf Q})$}
In Fig. \ref{CES_plots} we show the momentum dependence of the spin-correlation function ${\cal S}({\bf Q})$ averaged over an energy range of 5-10 meV. The data were acquired with an incident energy $E_i=50$ meV at temperatures 205 K, 170 K, and 5 K. The left column (a) (d), and (g) shows ${\cal S}({\bf Q})$ for the pseudo-hexagonal basal plane ${\bf Q}=(HK0)_H$. It is clear from the low temperature data (g) and (i) that there are three monoclinic domains in the hexagonal basal plane (each 60$^{\circ}$ apart). Their volume fractions of 56\%, 26\%, and 18\% respectively were determined from Fig.~\ref{CES_plots}(g) and duly incorporated into the forthcoming simulations. The unequal domain population can arise from symmetry breaking associated with the sample shape and the strain imposed by the sample mounting provisions. 

At $T=205$~K (Fig. \ref{CES_plots}(a), \ref{CES_plots}(b), and \ref{CES_plots}(c)) we observe diffuse scattering that we associate with short range correlated paramagnetic spin fluctuations. Consistent with rhombohedral symmetry, the six-fold structure of magnetic neutron scattering in the pseudo-hexagonal basal plane $L_H=0$, (Fig \ref{CES_plots}(a), \ref{CES_plots}(d), and \ref{CES_plots}(g)) becomes a three-fold structure for $L_H\ne 3\times$integer (Fig.~\ref{CES_plots}(c), \ref{CES_plots}(f), \ref{CES_plots}(i), and Fig. \ref{S_Q}). In these new data we see for the first time the direction, \mbox{(H 0 0)$_{M}$} (indicated by the green arrow \#3 in Fig.~\ref{CES_plots}(h)), along which coherent spin waves abruptly ``melt" into a broad paramagnon excitation at wave-vectors located between the acoustic spin wave branches. 

A dramatic change in the character of the magnetic excitations across the first order AFI to PI phase transition is also apparent in the ${\bf{Q}}$-$E$ slices of Fig.~\ref{Dispersions}(d). (Note: the qualitative features of Fig.~\ref{Dispersions}(d) persist up to at least 320~K as shown in Appendix~\ref{append_pure}, Fig.~\ref{disp_supp}(d)). The sharp excitations near 80~meV energy transfer along the $(-L~1~L)_M$ and $(-2~1~L)_M$ directions that survive to 320~K would appear to be phonons.  

\subsection{Exchange Interactions in the AFI Phase}

\subsubsection{Neutron Scattering}
\begin{table}[!t]
	\centering
	\begin{tabular*}{0.485\textwidth}{@{\extracolsep{\fill} } c|ccccc }
		\hline\hline\noalign{\smallskip}
		distance &$J_i$  & DFT & DFT & Data & sign of \\ 
		($\mathrm{\AA}$) & {($S$=1)} & V$_2$O$_3$ & (V$_{0.96}$Cr$_{0.04}$)$_2$O$_3$ & (Fit) &$\langle{\bf S}_i\cdot{\bf S}_j\rangle$\\
		[0.5ex] 
		\hline\noalign{\smallskip}
		2.75904 & $J_{\alpha}$ & {2.8(3)} &-1(2) & -6.0(2) &+  \\ 
		2.83083 & $J_{\beta 1}$ & {28.7(3)}&25(2) &27.7(2) &--  \\
		2.91789 & $J_{\beta 2}$  & {12.4(3)}& 9(2) & 7.7(2)&--  \\ 
		2.98538 & $J_{\gamma}$  & {-2.3(3)}& 3(2) & 0.0(2)&+  \\
		3.43336 & $J_{\epsilon}$ & {-3.9(5)}& -9(3)  & 2.0(2)&--  \\
		3.45420 & $J_{\delta}$ & {3.5(3)}& 4(2)  & 1.1(2)&--  \\
		3.63334 & $J_{\eta}$  & {0.6(3)}& 1(2)  & -2.0(2)&+ \\
		3.70177 & $J_{\zeta 1}$ & {2.5(2)}& 1(1)  & 7.1(2)&--  \\
		3.76876 & $J_{\zeta 2}$ & {-0.3(2)}& -1(1) & 7.1(2)&--  \\
		4.22293 & $J_{\theta}$  & {0.8(3)} & -4(2)  & 0&+ \\
		4.97765 & $J_{\iota}$   & {0.4(2)}& 3(1)  & 0&--\\
		5.00240 & $J_{\kappa}$  & {1.7(2)}& -1(1)  & 0&+ \\
		\hline\hline
	\end{tabular*}
	\caption{Magnetic exchange constants {(in units of meV)} for the AFI phase in (V$_{0.96}$Cr$_{0.04}$)$_2$O$_3$, obtained from both DFT calculations and direct fitting of the neutron data as described in the main text. Column 6 gives the sign of the indicated correlation $\langle{\bf S}_i\cdot{\bf S}_j\rangle$ for the AFI structure (Fig.~\ref{Phase_diagram}(c)). The consistent negative sign of the product between columns 5 and 6 indicates an unfrustrated magnet.}
	\label{LT_J}
\end{table}

\begin{figure}[!t]
	\includegraphics[width=1.01\columnwidth,clip]{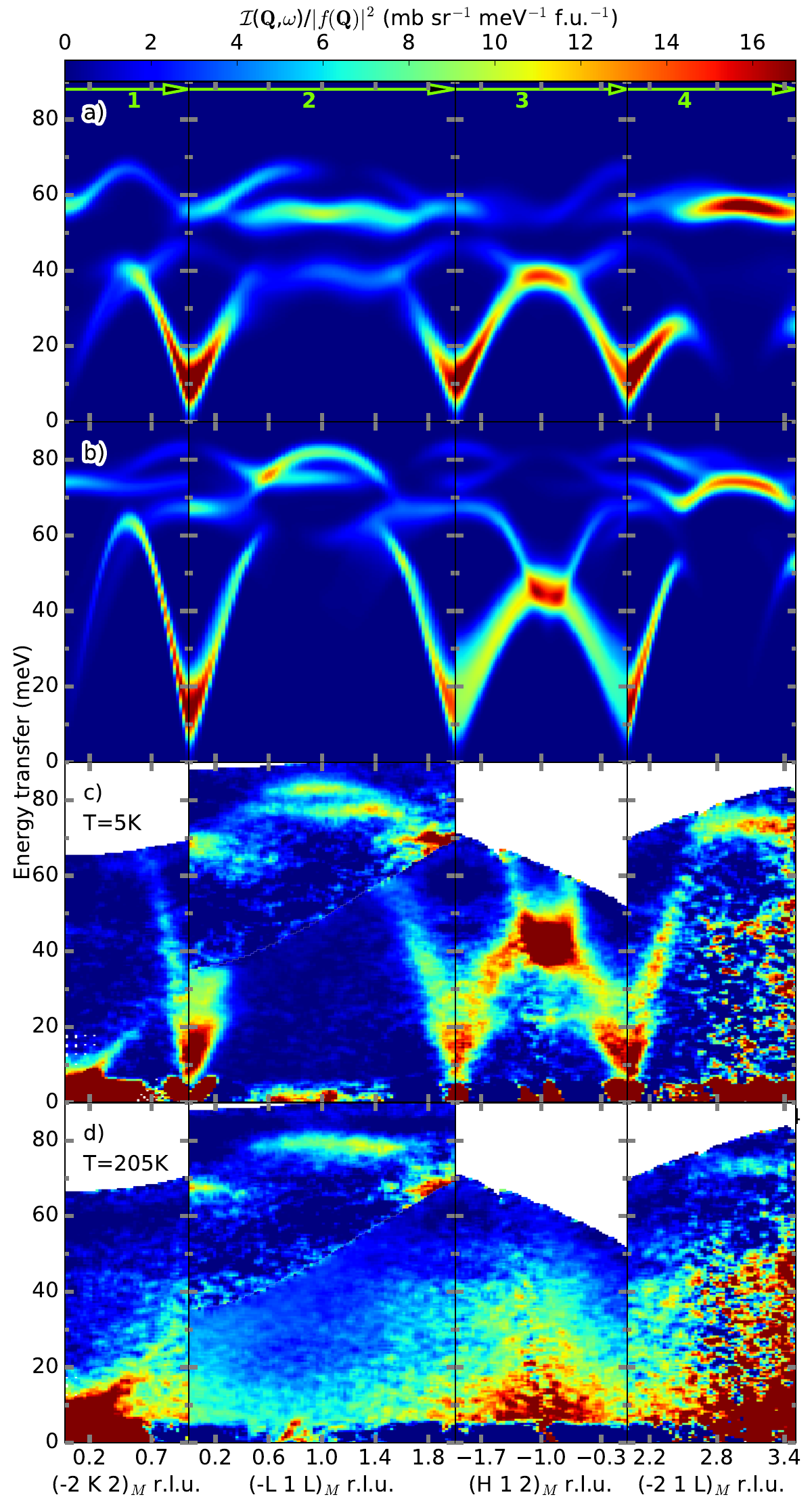}
	\caption{ \label{Dispersions} Neutron scattering intensity associated with spin waves along four high symmetry directions marked in Fig.~\ref{CES_plots} using (a) DFT calculated exchange constants and (b) exchange constants obtained from fits to the data. (c)(d) INS cross section measured at 5~K and 205~K ($E_i$~=~100 meV). Data from multiple Brillouin zones were assembled to cover the largest possible range of energy transfer. The data were divided by the squared magnetic form factor for presentation. Also, an incoherent background determined by azimuthal averaging of the same data was subtracted (Appendix~\ref{append_background}).} 
\end{figure}

Our comprehensive INS data set acquired for $E_i$~=~100~meV  provides access to both acoustic and optic spin wave excitations from the AFI (Fig.~\ref{Dispersions}(c)). Along all high-symmetry directions, sharp dispersive ridges of magnetic scattering are observed indicative of coherent spin-wave like excitations in the AFI. These can be described as the normal modes of excitation from the ordered state of an antiferromagnet described by a spin Hamiltonian of the form:
\begin{equation}
{\cal H}_{mag} =  \sum_{i,j}{J_{ij}  \textbf{S}_{i} \cdot \textbf{S}_{j}+\sum_{i}{D (S^z_{i})^2}}.
\label{hamiltonian}
\end{equation}
Here $J_{ij}$ are exchange constants for interaction between spins $i$ and $j$ ($S=1$) that we approximate as Heisenberg like (isotropic in spin-space). The symmetry inequivalent exchange interactions in the PI and AFI phases are defined in  Fig.~\ref{Phase_diagram}(b)(c). A derivation of the corresponding spin wave dispersion relation in the monoclinic AFI phase was first presented in Ref.~\cite{Bao_PRB_1998}. For simplicity, the measured 4.8 meV energy gap in the spin-wave excitation spectrum is wholly attributed to an uni-axial single ion anisotropy term $D$~=~0.13~meV, even though exchange anisotropy must also be present \cite{Word_1981_PRB,Bao_PRB_1998}.

For a given set of exchange constants, the inelastic neutron scattering cross section associated with spin wave excitations from the AFI state was calculated with linear spin wave theory (LSWT) using the SpinW program \cite{SpinW} and then convoluted with the energy dependent instrumental resolution function \cite{granroth2010sequoia,stone_review}. By comparing the measured and calculated spectra along the four high symmetry directions (Fig.~\ref{CES_plots}(g) and \ref{CES_plots}(h)), we inferred the best fit exchange parameters listed in column 5 of Table~\ref{LT_J}. (The error bars given are an upper bound expressing the range of each exchange parameter value that allows for an acceptable overall fit of the INS data in Fig. \ref{Dispersions}(c)). The INS that we associate with spin-waves (Fig.~\ref{Dispersions}(c)) is well accounted for by the model (Fig.~\ref{Dispersions}(b)), which gives confidence that the corresponding exchange Hamiltonian describes magnetic interactions within the AFI phase.  

\subsubsection{Density Functional Theory}

DFT electronic structure calculations for the monoclinic AFI were performed using the $T = 15$~K structure of Rozier {\it et al.} \cite{rozier2002comparative} with the full potential local orbital (FPLO) basis set \cite{PhysRevB.59.1743} and the GGA {and GGA+U} functional \cite{PhysRevLett.77.3865}. We obtained up to 12$^{th}$ near neighbor exchange constants by mapping total energy differences of various magnetic configurations to the Heisenberg model~\cite{jeschke2013}. Table~\ref{LT_J}, 3rd column displays the calculated exchange constants for pure V$_2$O$_3$ with $U$~=~3~eV. (see also Appendix~\ref{append_pure}). While these values were obtained from total energy calculations, where all contributions of crystal-field splittings and orbital hybridization paths --as dictated by the details of the crystal structure-- are integrally taken into account but difficult to extract individually, perturbation expansion considerations indicate that these exchange interactions strongly depend on crystal-field splittings and hoppings between a$_{1g}$ and e$_g$ states \cite{saha2009electronic} as previously found for other multi-orbital systems \cite{khomskii2014transition, Khaliullin_2005}. 

The experimental determination of the AFI phase exchange constants (Table~\ref{LT_J}, 5th column) conveniently serve to benchmark the \textit{ab initio} calculations of exchange interactions. While the agreement for the dominant exchange constants (3rd and 5th columns in Table~\ref{LT_J}) is rather good, however, we found that to achieve quantitative agreement with the experimental $J$ values, in particular the sign of $J_{\alpha}$, accounting for the 4\%~Cr ($S$~=~3/2) doping in the sample was necessary. Introducing Cr-doping into the DFT calculations is quite challenging and leads to significantly higher statistical error ranges. Two complimentary approaches were employed: (1) Replacing one V$^{3+}$ ion by a Cr$^{3+}$ ion in a large $\rm V_2O_3$ supercell where the different bonding environment of Cr$^{3+}$ is explicitly taken into account and (2) increasing the average nuclear charge from $Z$~=~23 (pure V$_2$O$_3$) to $Z$~=~23.04 ((V$_{0.96}$Cr$_{0.04}$)$_2$O$_3$) and using the virtual crystal approximation. Both of these approaches can deliver reasonable indications concerning the effect of Cr doping. We found that in both cases the Cr-doping indeed introduces a tendency towards FM  $J_{\alpha}<0$ in better agreement with the experimental data. As previously noted, local changes in the crystal structure induced by Cr-doping play an important role in the MIT \cite{lechermann2018uncovering}. However, the resulting exchange coupling constants from these calculations do not show drastic changes upon Cr-doping. An explanation for this is that the magnetic exchange constants are roughly $J \propto -t^2/(U+ \Delta)$ where $U \gg \Delta$ is the dominant scale and therefore they are not expected to be highly sensitive to effects related to changes of the order of the crystal field $\Delta$. The DFT calculated exchange constants for (V$_{0.96}$Cr$_{0.04}$)$_2$O$_3$ in the AFI phase with $U$~=~3~eV are shown in the 4th column of Table~\ref{LT_J}. Overall the in-plane interactions obtained from the spin-wave fit to the INS measurements are consistently accounted for by DFT while next-nearest neighbor exchange interactions linking neighboring pseudo-honeycomb lattice planes seem to be less well reproduced. 

The simulated ${\cal S}({\bf Q}, \omega)$ from these \textit{ab initio} exchange parameters using LSWT as implemented in SpinW is shown in Fig.~\ref{Dispersions}(a). Considering this is a first principles result, the similarity with measured INS in Fig.~\ref{Dispersions}(c) is rather good. Limitations of DFT and LSWT for this low spin ($S$~=~1) quantum magnet may contribute to discrepancies between the experimental data and theory.

\begin{table}
	\centering
	\begin{tabular*}{0.485\textwidth}{@{\extracolsep{\fill} } c|lccc }
		\hline\hline\noalign{\smallskip}
		distance   & &  & {DFT}  & DFT\\
		($\mathrm{\AA}$) &$J_j$ & $\equiv$ $J_i$ & {Pure $\rm V_2O_3$} &(V$_{0.96}$Cr$_{0.04}$)$_2$O$_3$ \\
 		[0.5ex] 
		\hline\noalign{\smallskip}
		2.71072 &$J_{1}$ &$J_{\alpha}$& {2.7(5)}  &-0.3(6) \\ 
		2.87799 &$J_{2}$ &$J_{\beta 1}$,$J_{\beta 2}$,$J_{\gamma}$& {9.3(2)} & 8.5(3) \\
		3.46255 &$J_{3}$ &$J_{\epsilon}$,$J_{\delta}$ & {2.1(2)} &  0.6(3)  \\ 
		3.68774 & $J_{4}$ &$J_{\eta}$,$J_{\zeta 1}$,$J_{\zeta 2}$& {0.7(1)} & 0.0(2) \\
		4.29734 & $J_{5}$ &$J_{\theta}$&  {-0.3(5)} & -1.2(7) \\
		4.94240 & $J_{6}$ &$J_{\iota}$,$J_{\kappa}$&  {1.9(1)} & 1.7(2) \\
		\hline\hline
	\end{tabular*}
	\caption{ DFT Calculated exchange constants {(in meV)} for the PI phase {of pure $\rm V_2O_3$ and} (V$_{0.96}$Cr$_{0.04}$)$_2$O$_3$. Column 3 shows the corresponding exchange constants for the lower symmetry AFI phase.}
	\label{HT_J}
\end{table}

\subsection{Exchange Interactions in the PI Phase}

Having shown that DFT-calculated exchange interactions in the AFI phase of (V$_{0.96}$Cr$_{0.04}$)$_2$O$_3$ provide a good account of the measured spin wave spectra, we use the same method to determine the exchange interactions in the PI phase, where the absence of spin wave excitations makes it harder to infer the exchange interactions from neutron scattering measurements. 

DFT total energy calculations for the rhombohedral PI were performed using the high temperature structure of Rozier {\it et al.} \cite{rozier2002comparative}. Table \ref{HT_J} shows the results for the first six exchange constants (identified by their bond distances) for pure V$_2$O$_3$ (column 4) and for V$_2$O$_3$ including Cr doping (column 5). The latter calculations were performed because, as mentioned earlier, we found this to be important particularly for the nearest neighbor $J_\alpha$ interaction in the AFI.  The corresponding exchange paths (bond vectors) are shown in Fig. \ref{Phase_diagram}(b). These exchange constants yield a Curie-Weiss temperature $\Theta_{CW}$ of $-400$~K, which is consistent with the experimental value of $\Theta_{CW}=-350$~K for (V$_{0.96}$Cr$_{0.04}$)$_2$O$_3$ ($\Theta_{CW}=-600$~K for pure V$_2$O$_3$) \cite{PhysRevB.2.3756}.

\begin{figure}
	\includegraphics[width=1.0\columnwidth,clip]{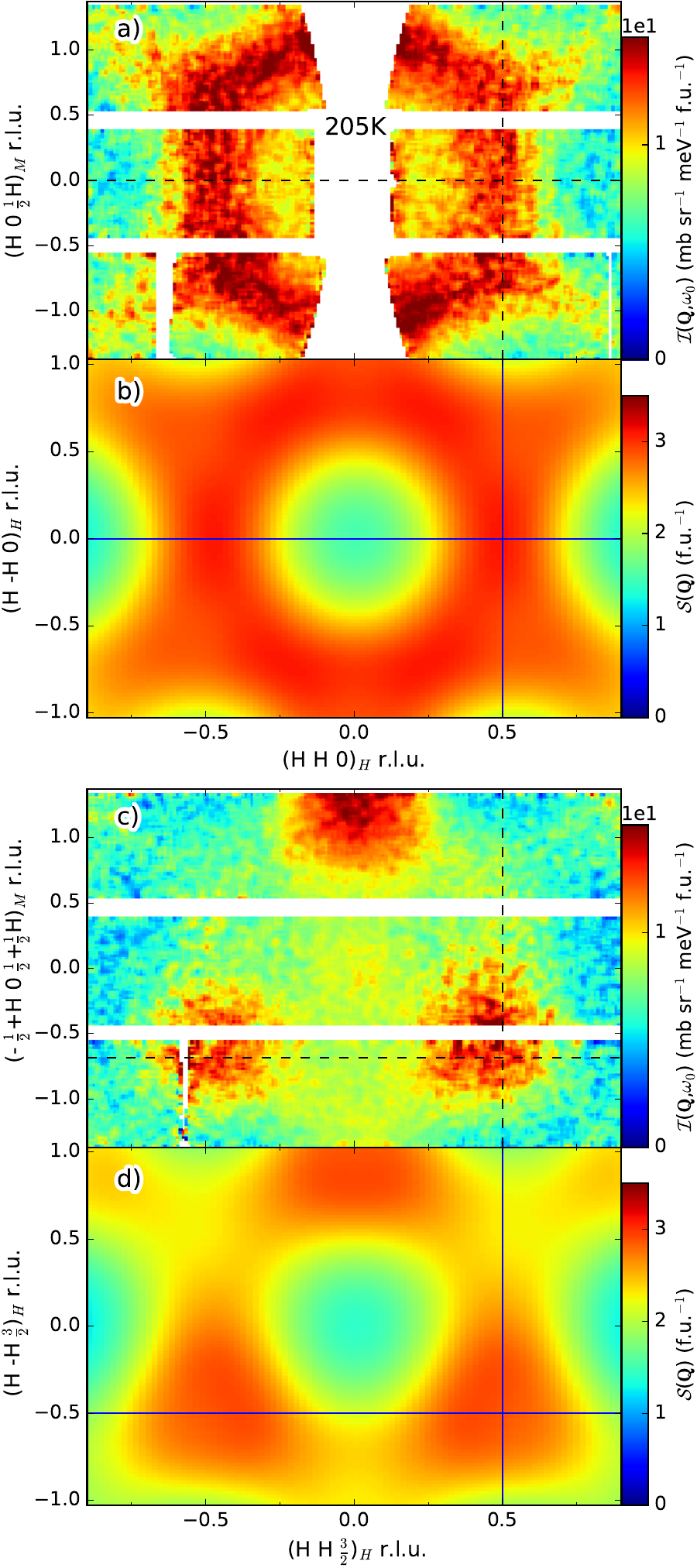}
	\caption{ \label{S_Q}(a)(c) Inelastic neutron scattering intensity from (V$_{0.96}$Cr$_{0.04}$)$_2$O$_3$ at 205~K at $\hbar\omega_0=7.5$~meV (averaging range = [5,10] meV) for wavevector transfer within pseudo-hexagonal basal planes (as in Fig. \ref{CES_plots}) at offsets along the pseudo-hexagonal c-axis of $L_H$~=~0 (a,b) and $L_H$~=~1.5 (c,d). (b)(d) These experimental data are compared with the calculated dynamic spin correlation function $\cal{S}({\bf Q})$ obtained with the complete set of DFT calculated exchange constants of the PI phase for (V$_{0.96}$Cr$_{0.04}$)$_2$O$_3$ given in Table~\ref{HT_J}. Lines indicate positions of cuts shown in Fig.~\ref{S_Q_cuts}.}
\end{figure}

\begin{figure}
	\includegraphics[width=1.0\columnwidth,clip]{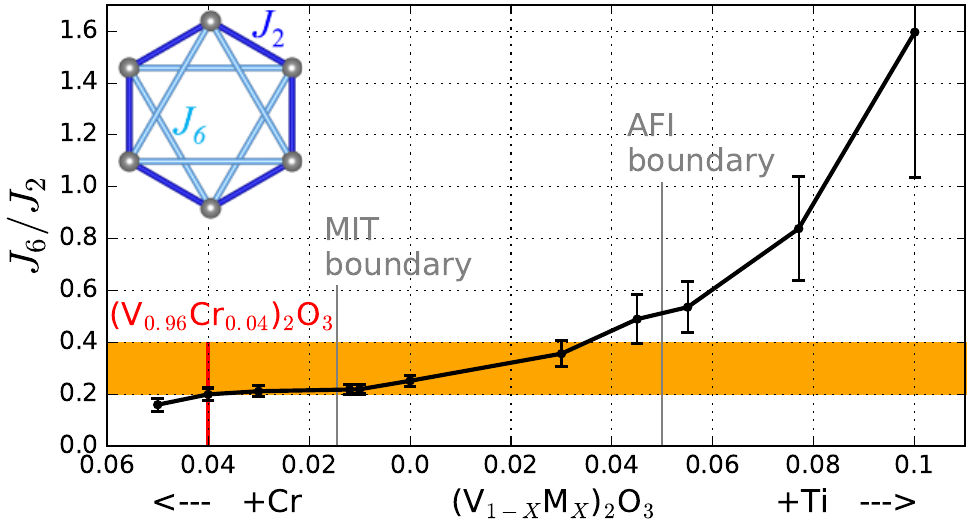}
	\caption{ \label{J2J6}  The ratio of the in-plane honeycomb lattice next-nearest and nearest neighbor exchange interaction values, $J_6$/$J_2$, throughout the phase diagram (Fig.~\ref{Phase_diagram}(a)) of V$_2$O$_3$ obtained from DFT calculations. The shaded orange region indicates the theoretically predicted range of frustrated interactions \cite{Albuquerque_PRB_2011}. It is clear that changes in the crystal structure induced from Cr-doping has only minor effects on $J_6$/$J_2$, as can also been seen in Table \ref{HT_J}.}
\end{figure}

\begin{figure}
	\includegraphics[width=1.0\columnwidth,clip]{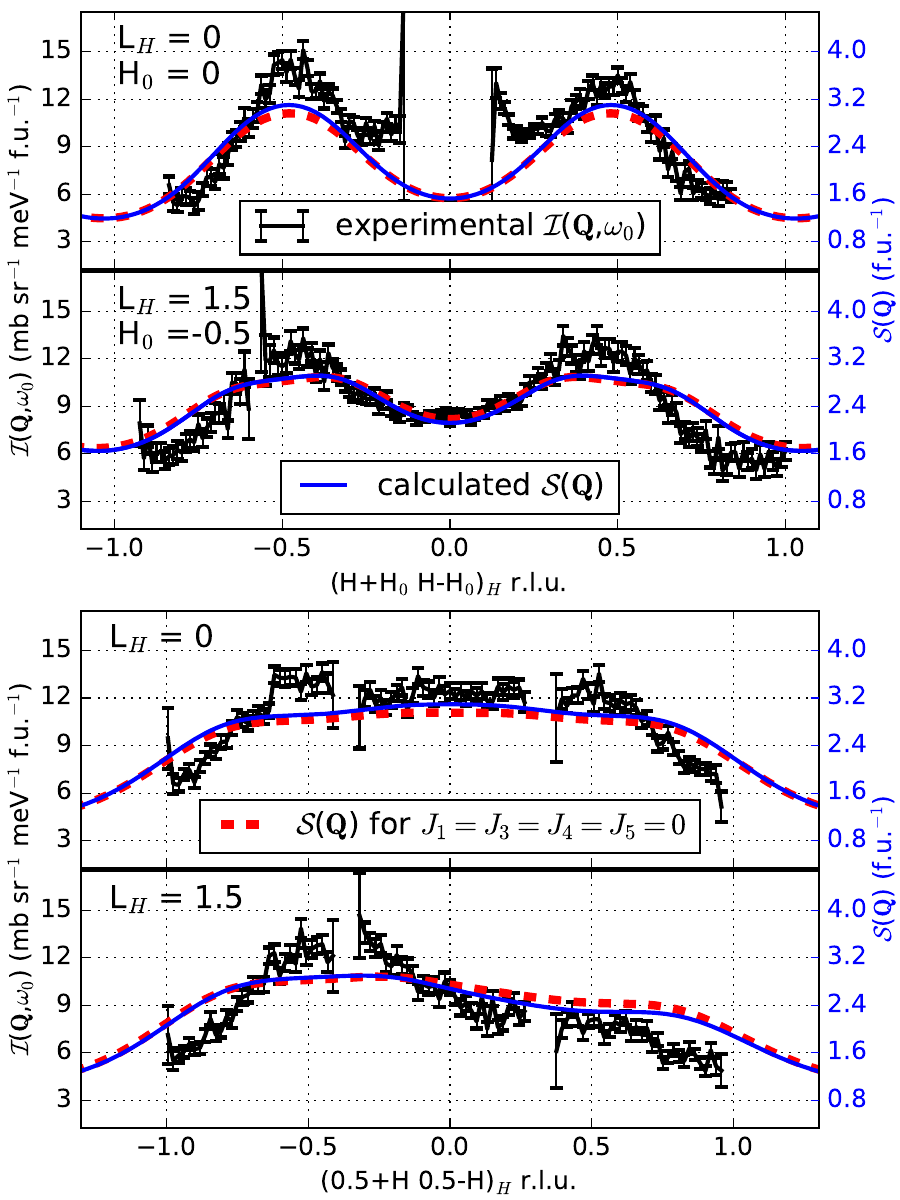}
	\caption{ \label{S_Q_cuts}  Cuts through the experimental  $\cal{I}$($\bf{Q}$,$\omega_0$) (black error bars) and the calculated $\cal{S}$($\bf{Q}$) (blue lines) for (V$_{0.96}$Cr$_{0.04}$)$_2$O$_3$. Top (bottom) panels show horizontal (vertical) cuts taken along the corresponding lines indicated in Fig. \ref{S_Q}. To show the importance of the dominant exchange constants $J_2=8.5(3)$ and $J_6=1.7(2)$, the red line shows the calculated $\cal{S}$($\bf{Q}$) when $J_1=J_3=J_4=J_5=0$.}
\end{figure}

The fact that the magnetic correlation length is on the atomic scale (Fig.~\ref{S_Q}) even for $T<\frac{1}{2}|\Theta_{CW}|$ indicates the PI phase is magnetically frustrated. Let us now examine whether or not the PI phase interactions inferred from DFT can account for the specific short ranged nature of the spin correlations. The self-consistent Gaussian approximation (SCGA) was previously shown to be effective for determining the equal time spin-correlation function, $\cal{S}(\mathbf Q)$, of geometrically frustrated magnets \cite{ConlonChalkerPRB}. Based on a spherical spin model \cite{spherical}, the softened spin configurations are weighted by the Boltzmann factor $e^{-\beta {E}}$ with
\begin{equation}
\beta {E} =  \frac{1}{2} \sum_{ij} (\beta \sum_n J_n A_{ij}^{(n)} + \lambda \delta_{ij}) s_i s_j.
\end{equation}
Here $s_i$ denotes one component of the spin vector $\mathbf S_i$ and $A^{(n)}$ is the adjacency matrix between $n^{th}$-nearest neighbors. The Lagrangian multiplier $\lambda$ is determined self-consistently to ensure the average spin length $\langle \mathbf S_i^2 \rangle = 1$. The rhombohedral structure of the PI is broken into six hexagonal sublattices and the Fourier transform is taken to obtain a quadratic interaction between sublattices. In reciprocal space, the condition on $\lambda$ is expressed as the trace of the inverse interaction matrix ($\sum_n J_n A^{(n)}$). The spin correlator as the two-body propagator is calculated at each desired $\mathbf Q$-point. We applied the SCGA method to determine spin correlations at $T=205$~K, where fluctuations of the classical spins are thermal. The calculated equal time spin correlation function $\cal{S}$($\bf{Q}$) captures the dominant features of the experimental INS data (see the matching broad features shown in Fig.~\ref{S_Q} and Fig.~\ref{S_Q_cuts}). 

\section{Analysis and Discussion}

\subsection{Frustrated Interactions in the PI Phase}

With the detailed set of exchange constants at hand (Table \ref{HT_J}) we are in a position to examine the nature of the frustrated PI phase. The two strongest exchange constants, $J_2$ and $J_6$ (see Fig.~\ref{Phase_diagram}(b) and Fig.~\ref{J2J6}), form quasi-two-dimensional honeycomb lattices with competing nearest neighbor ($J_2$) and next-nearest neighbor ($J_6$) interactions. As can be observed in  Fig.~\ref{S_Q_cuts}, $J_2$ and $J_6$ are the two indispensable exchange constants for generating the observed $\cal{S}$($\bf{Q}$) pattern, since the case of $J_1=J_3=J_4=J_5=0$ only slightly modifies the $\cal{S}$($\bf{Q}$) calculation. 

While $J_2$ favors bipartite AFM order where each spin is anti-parallel to its nearest neighbors, $J_6$ stabilizes a stripy phase \cite{Albuquerque_PRB_2011}. For (V$_{0.96}$Cr$_{0.04}$)$_2$O$_3$ the DFT calculations yield $J_6$/$J_2 \sim$~0.20 which place the PI phase in the valence bond crystal or spin liquid phase of the frustrated honeycomb model \cite{Albuquerque_PRB_2011,PhysRevB.85.060402}. The correlation length for such a model may be sufficiently short that the only significant inter-layer interactions (which are $J_5$, $J_3$, and $J_1$ in order of their strength) are rendered ineffective. Indeed, we note that the only out-of-plane exchange constant $J_5=-1.2(7)$ which is comparable in magnitude to $J_6$ does not substantially alter the pattern of magnetic excitations, and therefore does not appear to interfere with the frustration originating in the honeycomb basal plane. The importance of the next-nearest interaction $J_6$ within the honeycomb layers of pure V$_2$O$_3$ was previously noted based on GGA+U calculations \cite{PhysRevB.92.075121}. 

In order to estimate how widely this fundamental frustration mechanism is distributed over the PI phase, we again employed the virtual crystal approximation to perform DFT total energy calculations for different doping levels (with both Cr and Ti). The resulting $J_6$/$J_2$ values versus Cr and Ti doping are shown in Fig.~\ref{J2J6}. These results indicate that pure V$_2$O$_3$ lies near the optimal ratio of $J_6$/$J_2$ for frustrated interactions, and that this frustrated configuration remains present up to doping levels of $\sim$~4\% with either Cr or Ti. 

Examining the ${\bf{Q}}$-$E$ slices for the PI phase in Fig. \ref{Dispersions}(d) reveals factorization of the $\bf{Q}$ and $E$ dependence of the dynamic spin correlation function with a 40~meV bandwidth in energy that resembles the Weiss temperature and greatly exceeds $k_BT$. Comparison of Fig. \ref{Dispersions}(c) and \ref{Dispersions}(d) show this is not diffuse scattering of the AFI as it is weak at AF Bragg points. Instead such dynamic correlations resemble a number of strongly frustrated magnets that are described as quantum paramagnets such as $\rm ZnCr_2O_4$~\cite{ZCO_PRL} and $ \rm SrCr_{9p}Ga_{12-9p}O_{19}$~\cite{SCGO_p}.

The SCGA method breaks down at low temperatures when the interacting spin model develops magnetic order. A finite size scaling calculation (using all the LDA+U calculated exchange constants for (V$_{0.96}$Cr$_{0.04}$)$_2$O$_3$ in Table \ref{HT_J}) estimates the ordering temperature to be $\sim$10~K. Comparison to the Curie-Weiss temperature of $\sim$400~K yields a frustration index of $f=40$, indicating a high degree of magnetic frustration in the PI phase. The $T \approx$~10~K phase transition anticipated for the PI spin Hamiltonian is however preempted in (V$_{0.96}$Cr$_{0.04}$)$_2$O$_3$ at $T_N=185$~K by the first order magneto-structural transition that relieves magnetic frustration and gains magnetic exchange energy at the expense of lattice strain energy. DFT calculations and neutron scattering measurements indicate that the magnetic exchange interactions are significantly modified at this transition (compare Tables \ref{LT_J} and \ref{HT_J}). In particular, the three identical nearest neighbor interactions within the PI honeycomb lattices (DFT: $J_2=8.5(3)$~meV) split into three distinct interactions in the AFI phase (DFT[Neutron]: $J_{\beta_1}=25(2)[27.7(2)]$~meV, $J_{\beta_2}=9(2)[7.7(2)]$~meV, and $J_\gamma=3(2)[0.0(2)]$~meV). This relieves frustration and gives way to the stripy ordered phase favored by the next-nearest neighbor interaction (DFT: $J_6=1.7(2)$~meV).  

\subsection{Unfrustrated Interactions in the AFI Phase}
Referring to the experimental exchange constants in the 5th column of Table \ref{LT_J}, we note they are consistently {\em un-}frustrated: Spins that interact antiferromagnetically (ferromagnetically) are also antiferromagnetically (ferromagnetically) correlated in the ordered state (see column 6 of Table \ref{LT_J}). Thus, while exchange interactions in the PI phase realize a frustrated honeycomb antiferromagnet with competing nearest ($J_2$) and next-nearest neighbor ($J_6$) antiferromagnetic interactions, the modified nearest neighbor interactions in the AFI phase are all simultaneously satisfied in the long range ordered magnetic state (i.e., the signs of the 5th and 6th columns of Table \ref{LT_J} are consistently opposite). 

The raw experimental evidence for unfrustrated interactions is the large bandwidth of magnetic excitations in the AFI phase (80~meV). Because the bandwidth exceeds $k_BT_N$ by a factor of four, the collapse of the ordered state upon heating occurs before significant thermal population of spin waves. Correspondingly, there is no build up of magnetic correlations upon cooling the frustrated PI towards the  $T_N=185$~K phase transition. Comparison of Fig.~\ref{Dispersions}(c) and Fig.~\ref{Dispersions}(d) shows a clear upward shift of magnetic spectral weight, which indicates lowering of the magnetic exchange energy \cite{ZCO_PRL} as frustration is relieved at the first order PI to AFI transition. 

The nearest neighbor inter-plane interaction between face-sharing vanadium atoms, $J_\alpha=-6.0(2)$~meV, is the main FM interaction that favors FM sheets in the AFI. Our results confirm that $J_{\beta}$ and $J_{\zeta}$ are stronger than $J_{\alpha}$, and therefore suggest superexchange via V-O-V paths contribute to the realization of a strong ferromagnetic alignment for the vertical V pairs. The values of $J_{\alpha}=-4.1$~meV and $J_{\beta 1}=18.4$~meV determined by the LDA+$U$ study of \cite{Ezhov_1999} are consistent with these values. Further, the authors' suggestion \cite{Ezhov_1999} that the next-nearest neighbor exchange interactions are significant agrees with our result of $J_{\zeta}=7.1(2)$~meV.

It is interesting to compare the experimental and \textit{ab initio} exchange interactions for (V$_{0.96}$Cr$_{0.04}$)$_2$O$_3$ with those of pure Cr$_2$O$_3$. Samuelsen {\it et al.} \cite{samuelsen1970inelastic} showed that only the first two exchange interactions are significant in Cr$_2$O$_3$, with the nearest neighbor interaction dominant at 7.5~meV. Thus the range of interactions in (V$_{0.96}$Cr$_{0.04}$)$_2$O$_3$ is considerably longer than in Cr$_2$O$_3$, which is broadly consistent with greater 3$d$-electron delocalization near the Mott transition of V$_2$O$_3$.

In addition to indicating a transition from a frustrated to an un-frustrated magnet, the doubling of the magnetic bandwidth from 40~meV to 80~meV during the PI-AFI transition serves as another sign of its first order nature \cite{frandsen2016volume}. Tables \ref{LT_J} and \ref{HT_J} show the magnetic exchange interactions are tightly coupled with the crystal structure and lattice parameters. Therefore, spin-phonon coupling is expected in V$_2$O$_3$ and is indeed evidenced in Brillouin scattering studies \cite{Brillouin_scattering_study_2006}. Comparison of  Fig.~\ref{Dispersions}(c) and \ref{Dispersions}(d) shows phonon excitations near 70-80~meV which appear to overlap with optical magnon branches. The intensity enhancement of the magnon modes in this region compared with the LSWT structure factor shown in \ref{Dispersions}(b) could be a result of magnon-phonon coupling. It is also likely that the overall dispersion of the phonon normal modes is significantly affected by both spin-phonon and magnon-phonon interactions. To quantify these effects as well as the relationship between the magnetic interactions and the PI-AFI phase transition, it will be interesting future work to calculate quantitative values for the magneto-elastic coupling strengths. Magnetoelastic effects near room temperature present great opportunities for applications. Strain in V$_2$O$_3$ thin films has for example been shown to influence the metal-insulator transition \cite{Strain_MBE_V2O3,Strain_PLD_V2O3} and the coercive field of nickel thin films deposited on V$_2$O$_3$ is almost doubled at the MIT \cite{APL_bilayer1, APL_bilayer2}.

\section{Conclusion}
In summary, magnetic exchange interactions in the AFI phase of (V$_{0.96}$Cr$_{0.04}$)$_2$O$_3$ were determined with INS measurements of both acoustic and optic spin waves throughout the monoclinic Brillouin zone. Ordered by magnitude the leading interactions are $J_{\beta}$ and $J_{\zeta}$, which are both antiferromagnetic, and the nearest-neighbor interlayer interaction $J_{\alpha}$, which is ferromagnetic (Fig.~\ref{Phase_diagram}(c)). These and indeed all interactions in the AFI phase are satisfied within the observed low $T$ magnetic structure. In the PI phase diffuse inelastic magnetic neutron scattering and DFT calculations combined with a Gaussian approximation to the spin correlation function show this is a quasi-two-dimensional cooperative paramagnet with frustrated nearest ($J_2$) and next-nearest neighbor ($J_6$) interactions within the puckered honeycomb layers that form the corundum structure (Fig.~\ref{Phase_diagram}(b)). We infer that the strong suppression of magnetic order that results from this frustration is key to preventing the AFI phase from engulfing the Mott-like PM to PI phase boundary in $\rm V_2O_3$.

The insulating states of compounds with an experimentally accessible Mott-like phase boundary including $\rm Ni(S_{1-x}Se_x)_2$,\cite{PhysRevB.68.094409} GaTa$_4$Se$_8$ \cite{kim2014spin,jeong2017direct}, Na$_4$Ir$_3$O$_8$ \cite{PhysRevLett.99.137207,PhysRevLett.102.186401}, and the quasi-two-dimensional organic systems $\rm \kappa-(ET)_2 Cu[N(CN)_2]Cl$, $\rm \kappa-(ET)_2 Cu_2 (CN)_3$ \cite{Senthil_2008}, and $\rm EtMe_3 Sb[Pd(dmit)_2 ]_2$ \cite{kanoda_mott} have all been found to be magnetically frustrated. Conversely the spin-liquid-like properties of the latter two compounds have been associated with the increased range and nature of spin-spin interactions near the Mott transition. The present work on $\rm (V_{1-x}Cr_x)_2O_3$ reinforces these interesting links between frustrated magnetism and the Mott transition, and signifies that the ongoing searches for a Mott insulator with a paramagnetic MIT is tied to the search for spin liquids \cite{QSL_review,pustogow2018quantum}. 
  
\section{Acknowledgements}
The authors would like to acknowledge helpful discussions with Sandor Toth, Arnab Banerjee, Michael Lawler, and Oleg Tchernyshyov. R.V. thanks Frank Lechermann for useful discussions. This project was supported by UT-Battelle LDRD \#3211-2440. Work at IQM was supported by the U.S. Department of Energy, Office of Basic Energy Sciences, Division of Materials Sciences and Engineering through Grant No. DE-SC0019331. This research used resources at the Spallation Neutron Source, a DOE Office of Science User Facility operated by the Oak Ridge National Laboratory (ORNL). C.B. was supported through the Gordon and Betty Moore foundation under the EPIQS program GBMF4532. This research used resources of the National Energy Research Scientific Computing Center, a DOE Office of Science User Facility supported by the Office of Science of the U.S. Department of Energy under Contract No. DE-AC02-05CH11231. Work of J.H. and O.D. at ORNL was supported by the U.S. Department of Energy, Office of Basic Energy Sciences, Division of Materials Sciences and Engineering through Grant No. DE-SC0016166. The work at BIT was supported by the National Science Foundation of China Grant No. 11572040. W.B. was supported by National Natural Science Foundation of China (Grant No. 11227906). The work in Frankfurt was supported by the Deutsche Forschungsgemeinschaft under Grant SFB/TRR 49.


\appendix

\begin{figure}[!h]
	\centering
	\includegraphics[width=1.02\linewidth]{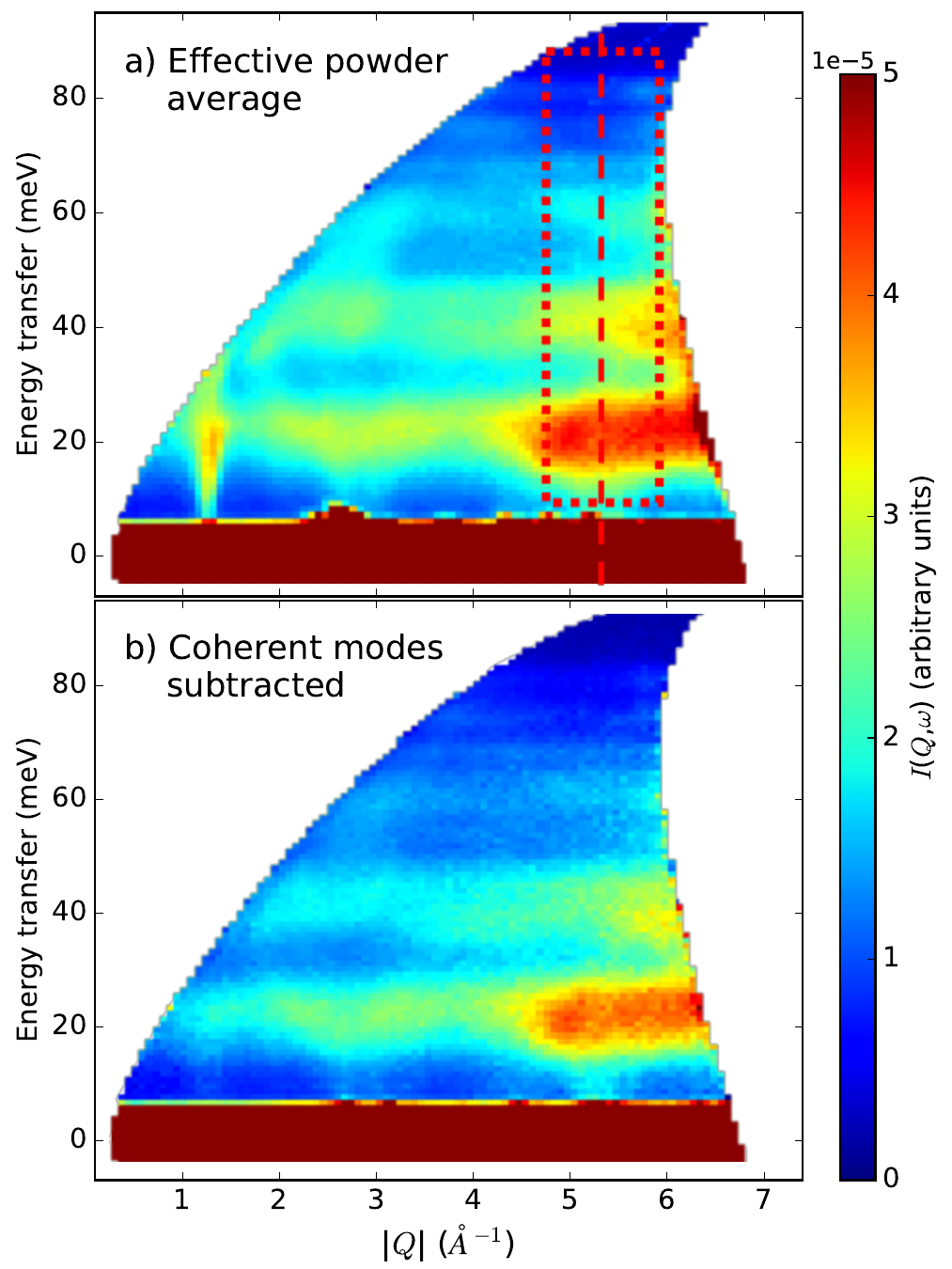}
	\caption{(a) An effective powder average (taken over all collected sample rotation angles) for the single crystal (V$_{0.96}$Cr$_{0.04}$)$_2$O$_3$ $T$~=~5~K data. The dashed red lines indicate the range of wave vector transfer used to generate the phonon DOS shown in Fig.~\ref{phonon_DOS}. (b) Powder average of the same data excluding scattering that depends on the azimuthal angle, $\phi$. Such filtered data were utilized as background.}
	\label{powder_background}
\end{figure}
\section{Data Processing Methods}
\subsection{Incoherent background subtraction}\label{append_background}

With the new capabilities of the Mantid Data Analysis Software \cite{arnold2014mantid}, it is possible to subtract a $\phi$ independent ($\phi$ being the azimuthal rotation angle of ${\bf k}_f$ around ${\bf k}_i$) background signal from the 4D $\cal{S}$(${\bf{Q}}$,$\omega$) data. This is done by taking a judiciously sampled powder average of the $\cal{S}$(${\bf{Q}}$,$\omega$) including only incoherent $\phi$-independent components and excluding coherent $\phi-$dependent components. (See Fig.~\ref{powder_background}). Once this special powder average is obtained it can readily be subtracted from the $\cal{S}$(${\bf{Q}}$,$\omega$) data as a background. 

\subsection{Phonon density of states (DOS) calculations}
\label{phonorm}

\begin{figure}
	\includegraphics[width=1.03\linewidth]{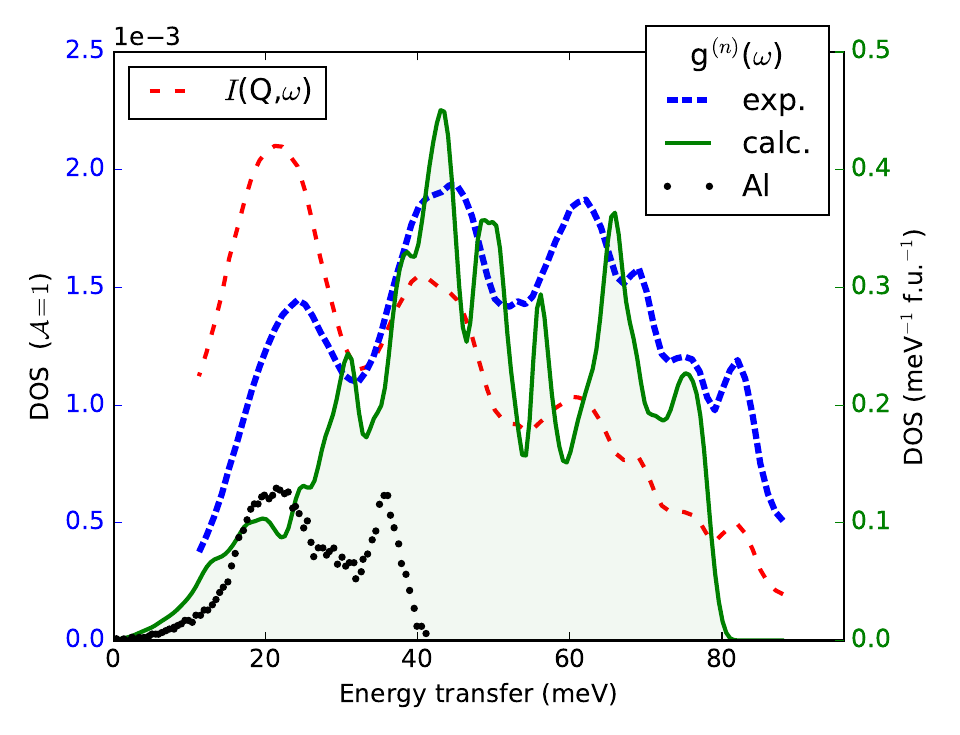}
	\caption{ \label{phonon_DOS} Phonon DOS $g^{(n)}(\omega)$ for (V$_{0.96}$Cr$_{0.04}$)$_2$O$_3$ extracted from the effective powder average of the INS data (See Fig. \ref{powder_background}(a)). This is compared with theoretical phonon DOS calculated for V$_2$O$_3$ in the low-temperature AFI phase. Also shown is the phonon DOS for aluminum from Ref. \cite{PhysRevB.77.024301}. A contribution to the experimental data from aluminum is inevitable due to the aluminum sample mount and cryostat windows.  } 
\end{figure}

Normalization of the scattering data was achieved through the incoherent phonon scattering cross section. The neutron weighted phonon DOS $g^{(n)}(\omega)$ was theoretically calculated with VASP + Phonopy for the AFI phase of V$_2$O$_3$. This is shown as a solid green line in Fig. \ref{phonon_DOS} in units of meV$^{-1}$f.u.$^{-1}$ (5 atoms per formula unit).

The experimental phonon DOS was obtained by first taking a constant-$Q$ cut through the effective powder average (generated by averaging over all collected sample rotation angles) of the $T$~=~5~K INS spectra. In this case the cut is taken by integrating in a rectangular area at high momentum transfer centered around $|Q|$~=~5.3~$\rm \AA^{-1}$, as shown by the dashed red lines in Fig. \ref{powder_background}(a). The result of this integration is plotted as the dashed red line (not to scale) in Fig.~\ref{phonon_DOS}. This cut from the experimental data can then be converted into the neutron weighted generalized phonon DOS $g^{(n)}(\omega)$ (the blue line in Fig. \ref{phonon_DOS}) via the following formula:
\begin{equation}
g^{(n)}(\omega)=\frac{{\cal A}I({Q},\omega)}{ \frac{\sigma_{\rm V_2O_3}}{4\pi} \left( \frac{(\hbar Q)^2}{2M}/\hbar\omega \right)(n(\omega,T)+1)}
\end{equation}
where $I({Q},\omega)$ is the proton charge normalized detector counts binned in ${Q=|{\bf Q}|}$ and $\omega$, $n(\omega,T) = [e^{\beta \omega}-1]^{-1}$, $\sigma_{\rm V_2O_3}~=~22.88$~barn/f.u. is the total scattering cross section (coherent plus incoherent) per formula unit for the sample, and $M$ is the mass per formula unit. ${\cal A}$ is the normalization constant, which was adjusted to achieve the best overlap between the experimental and theoretical $g^{(n)}(\omega)$ traces in Fig.~\ref{phonon_DOS}. This procedure yields ${\cal A}=200(50)$~barn$\times$coulomb/meV/counts. Using this normalization factor the conversion of raw scattering intensities to a normalized cross section is as follows: ${\cal I}({\bf Q},\omega)={\cal A}I({\bf Q},\omega)$.

Some of the discrepancy between the experimental and theoretical $g^{(n)}(\omega)$ may arise from contributions to scattering intensity from the aluminum can sample holder. Previously measured phonon DOS for aluminum \cite{PhysRevB.77.024301} shows  its strongest  peak is near 20 meV energy transfer. This is consistent with the extra intensity seen near 20 meV in the experimental $g^{(n)}(\omega)$ when comparing with the theoretical $g^{(n)}(\omega)$ for V$_2$O$_3$. We estimate this normalization procedure is accurate to within 25\%.

\begin{figure}[!h]
	\includegraphics[width=1.01\columnwidth,clip]{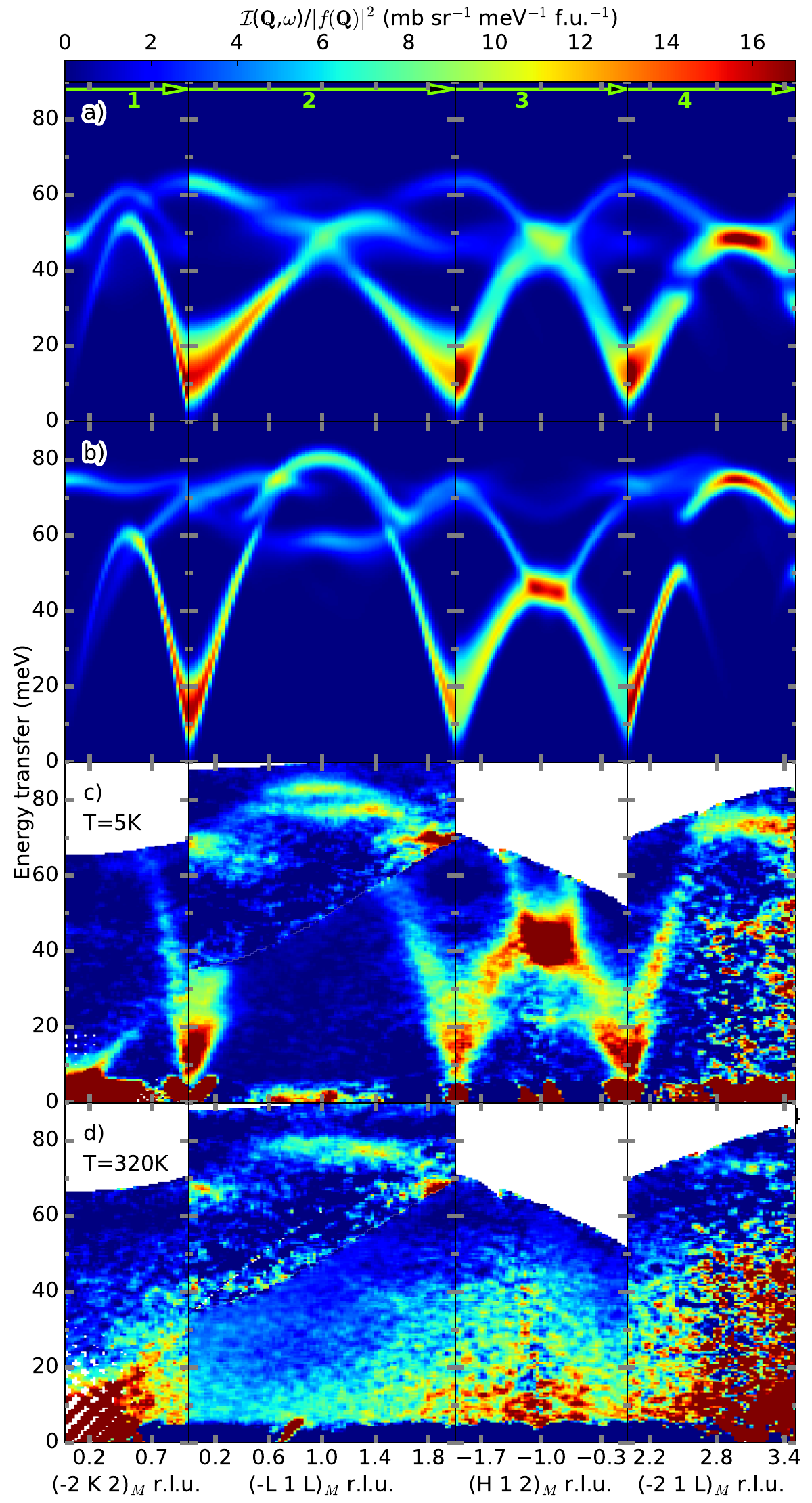}
	\caption{ \label{disp_supp} (a)(b) Neutron scattering intensity associated with spin waves along four high symmetry directions marked in Fig.~\ref{CES_plots} calculated with SpinW \cite{SpinW} for exchange constants shown in Table \ref{LT_J_pure_sample}. (c)(d) INS cross section measured at 5~K and 320~K ($E_i$~=~100~meV). Data from multiple Brillouin zones were assembled to cover the largest possible range of energy transfer. As before, the data were divided by the squared magnetic form factor for presentation and the incoherent background was subtracted from the data (Appendix~\ref{append_background}).} 
\end{figure}

\begin{table}
	\centering
	\begin{tabular*}{0.485\textwidth}{@{\extracolsep{\fill} } cccc }
		\hline\hline\noalign{\smallskip}
		$J_i$ & distance ($\mathrm{\AA}$) & Pure V$_2$O$_3$ & Fig. \ref{disp_supp}(b) \\ [0.5ex] 
		\hline\noalign{\smallskip}
		$J_{\alpha}$ & 2.75904 & 2.8(3) & -2.7 \\ 
		$J_{\beta 1}$ & 2.83083 & 28.7(3) & 27.5 \\
		$J_{\beta 2}$ & 2.91789 & 12.4(3) & 7.7 \\ 
		$J_{\gamma}$ & 2.98538 & -2.3(3) & 2.8 \\
		$J_{\epsilon}$ & 3.43336 & -3.9(5) & -6.1 \\
		$J_{\delta}$ & 3.45420 & 3.5(3) & 1.9 \\
		$J_{\eta}$ & 3.63334 & 0.6(3) & -1.0 \\
		$J_{\zeta 1}$ & 3.70177 & 2.5(2) & 2.5 \\
		$J_{\zeta 2}$ & 3.76876 & -0.3(2) & {\bf{6.5}} \\
		$J_{\theta}$ & 4.22293 & 0.8(3) & -1.6 \\
		$J_{\iota}$ & 4.97765 & 0.4(2) & 2.0 \\
		$J_{\kappa}$ & 5.00240 & 1.7(2) & 0.0 \\
		\hline\hline
	\end{tabular*}
	\caption{Exchange constants for monoclinic pure V$_2$O$_3$ (in~meV) calculated with DFT using two unit cells as explained in the text below. The resultant INS cross sections from the spin waves with these exchange constants are shown in Fig. \ref{disp_supp}(a). Also, shown in Fig. \ref{disp_supp}(b) is the LSWT result using the set of DFT calculated exchange parameters from Table \ref{LT_J} for (V$_{0.96}$Cr$_{0.04}$)$_2$O$_3$ with adjustments made (as described in the text) within statistical error tolerances, except for $J_{\zeta 2}$ (shown in bold). $J_{\zeta 2}$ must be increased to 6.5 meV which is outside of the statistical error bar interval to achieve the level of agreement with INS reflected by Fig. \ref{disp_supp}(b). }
	\label{LT_J_pure_sample}
\end{table}

\section{Extended Data Modeling}

\subsection{DFT calculated exchange constants}\label{append_pure}
The initial DFT calculations of exchange interactions were performed for pure monoclinic V$_2$O$_3$. These yielded results with smaller statistical errors than when including the effects of chromium doping. As mentioned in the main text, these calculations were performed with the full potential local orbital (FPLO) basis set \cite{PhysRevB.59.1743} and the GGA functional \cite{PhysRevLett.77.3865}. Total energies for large sets of different spin configurations were calculated with GGA+$U$ using the atomic limit double counting correction \cite{PhysRevB.79.035103}. We fix $J_H$~=~0.68~eV \cite{PhysRevB.54.5368} and vary $U$. For the LSWT modeling, we only employ the $U$~=~3~eV values. 

The calculation was carried out to obtain the first 12 exchange constants of monoclinic pure V$_2$O$_3$ by combining the total energies of two supercells: (1) A $2\times2\times2$ supercell with P$\overline{1}$ symmetry, which leaves 8 V$^{3+}$ ions inequivalent (2) A $1\times\sqrt{2}\times\sqrt{2}$ with P$\overline{1}$ symmetry, also containing 8 inequivalent V$^{3+}$ ions. While neither of the two structures allow for resolution of all 12 exchange constants, the combined equations have enough information. 

The exchange constants resulting from this calculation at $U$ = 3 eV are listed in the 3rd column of Table~\ref{LT_J_pure_sample}. The equations determined for the  spin configurations analyzed indicate some correlations between the inferred exchange constants, in particular between $J_{\alpha}$ and $J_{\epsilon}$. Statistical errors are on the order of 0.5 meV, providing a reasonable degree of confidence for all except the smallest exchange constants, $J_{\zeta 2}$ and $J_{\iota}$. The resultant neutron scattering intensity obtained from these exchange constants through LSWT as implemented in SpinW \cite{SpinW} are plotted in Fig. \ref{disp_supp}(a). 

We now return to the DFT calculated exchange constants for (V$_{0.96}$Cr$_{0.04}$)$_2$O$_3$ listed in Table~\ref{LT_J}. As the statistical error in these  is around 2 meV, liberty may be taken to make adjustments within these error ranges to bring the calculated spin wave scattering into agreement with observations. Such optimally adjusted $J$ values are shown in column 4 of Table~\ref{LT_J_pure_sample}. The resultant neutron intensity is plotted in Fig.~\ref{disp_supp}(b). The only adjustment that must stray beyond the tolerances of the statistical errors from Table~\ref{LT_J} in order to achieve consistency with experiment is the value of $J_{\zeta 2}$ (set as 6.5~meV, shown in bold). The DFT methods may have trouble accounting for this particular interaction due to accuracy limitations of the structural model for such long-range superexchange interactions.  Nevertheless, this alternate set of exchange parameters yields a pattern of scattering that is consistent with the neutron data, {which indicates considerable correlated uncertainty in deriving exchange constants from scattering data.} Comparison of column 5 in Table~\ref{LT_J} and column 4 of Table~\ref{LT_J_pure_sample} indicate the experimental error bar in the determination of these exchange constants from neutron scattering data. We note that in the original fitting, we set $J_{\theta} = J_{\iota} = J_{\kappa} = 0$ to reduce the number of fitting parameters. Clearly, with the use of 12 exchange constants to describe the neutron data (as with the ab initio DFT calculations) instead of 9 for our standard direct fitting, we may expect increased correlated uncertainties in the extracted exchange constants. Nonetheless, the overall qualitative trends and magnitudes of the two presented sets of exchange constants which fit the data are similar, thus preserving the validity of the associated discussion in the main text. 

\begin{figure}
	\includegraphics[width=1.0\columnwidth,clip]{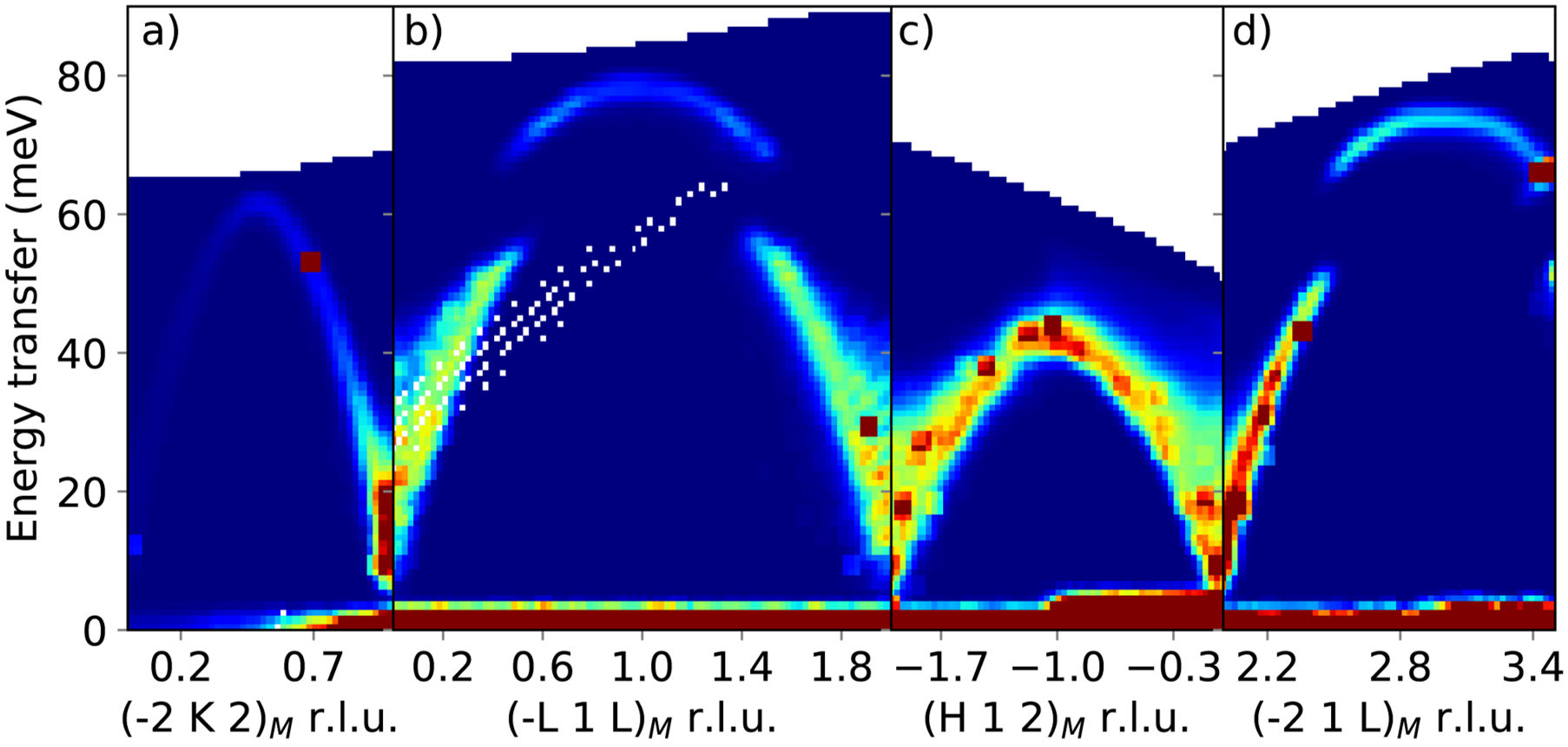}
	\caption{ \label{McVine} MCViNE simulation of the same ${\cal I}({\bf Q},\omega)$ slices shown in Fig. \ref{Dispersions}(c) and Fig. \ref{disp_supp}(c). This simulation takes into account all instrumental effects and shows similar linewidth broadening as in the experimental scattering data.}
\end{figure}
\begin{figure}
	\includegraphics[width=1.025\columnwidth,clip]{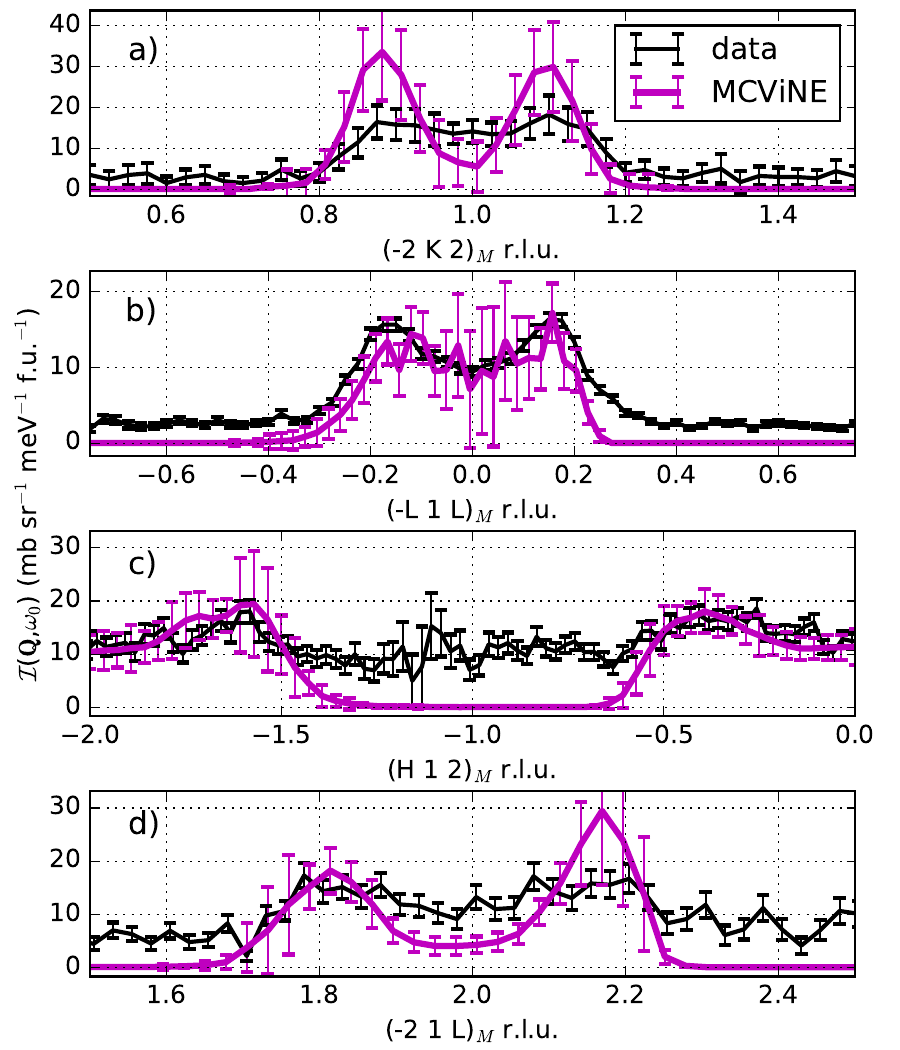}
	\caption{ \label{McVine_cuts} Constant energy cuts at $\hbar\omega_0$~=~30 meV (averaging range = [29,31] meV) through the experimental data for (V$_{0.96}$Cr$_{0.04}$)$_2$O$_3$ at 5~K (Fig.~\ref{Dispersions}(c)) and the corresponding MCViNE simulation (Fig.~\ref{McVine}). } 
\end{figure}

\subsection{Resolution effects}
Looking at Fig.~\ref{Dispersions}(a)-\ref{Dispersions}(c) as well as Fig.~\ref{disp_supp}(a)-\ref{disp_supp}(c), more broadening is noticeable in the experimental data (c) of the acoustic magnon branches than is predicted by the simulations (a)(b) with LSWT via SpinW. This discrepancy is present despite using consistent averaging ranges to produce each slice and the nominal energy resolution of SEQUOIA to approximate resolution effects within SpinW. To determine whether this reflects the physics of $\rm V_2O_3$ or more intricate ${\bf Q}$-resolution effects, we performed Monte Carlo ray-tracing simulations of the experiment using MCViNE~\cite{lin2016mcvine}. 

The simulation follows the routine MCViNE simulation procedure \cite{lin2016mcvine} that involves four steps. First the neutron beam $\sim$12 cm upstream of the sample position was simulated. In the second step, the simulated beam was scattered by a virtual sample, which is a plate of 4.6 cm x 4.6 cm x .57 cm and has a scattering kernel of a dispersion surface (see the supplemental materials of Ref.~\cite{lin2016mcvine}) specified by an analytical dispersion function. The virtual sample is aligned as in the experiment and the goniometer angle was swept from -90$^{\circ}$ to 90$^{\circ}$ in 2$^{\circ}$ steps. In the third step, the interception of simulated scattered neutrons by the SEQUOIA~\cite{granroth2010sequoia} detector system were simulated and a series of NeXus files \cite{Konnecke_neXus} were generated. In the last step, the NeXus files were reduced and slices were taken as for the real neutron scattering data. 

The results are shown in Fig.~\ref{McVine}. By examining constant energy cuts as shown in Fig.~\ref{McVine_cuts}, one sees that the broadening of the acoustic spin waves in the actual neutron data is for the most part reproduced by this simulation of realistic instrumental effects. There is however, a tendency towards sharper peaks in the simulation (especially in Fig.~\ref{McVine_cuts}(a) and \ref{McVine_cuts}(d)). This indicates some spin-wave damping, which is not unexpected for a quantum magnet with quenched disorder (due to chromium doping) near the MIT. But overall, the coherent magnon approximation provides a reasonable description of the full data set.

\bibliography{v2o3_ref}

\end{document}